\documentclass{moriond}

\def\Journal#1#2#3#4{{#1} {\bf #2}, #3 (#4)}

\def\NPB{{\em Nucl. Phys.} B}

\def\EPJC{{\em Eur. Phys. Journal}  C}

\def\PRL{\em Phys. Rev. Lett.}

\def\PRC{{\em Phys. Rev.} C}

\def\JINST{{\em JINST}}

\usepackage{amsmath}
\def\be{\begin{equation}}
\def\ee{\end{equation}}
\def\bea{\begin{eqnarray}}
\def\eea{\end{eqnarray}}

\newcommand{\Photo}

\begin{document}
\vspace*{4cm}
\title{Final results from the CUPID-Mo $0\nu\beta\beta$ experiment}

\author{ T. Dixon on behalf of the CUPID-Mo collaboration }

\address{Université Paris-Saclay, CNRS/IN2P3, IJCLab, 91405 Orsay, France}

\maketitle\abstracts{
CUPID-Mo is a demonstrator experiment for the next generation $0\nu\beta\beta$ experiment CUPID. It consisted of an array of 20 Lithium Molybdate (LMO) cryogenic calorimeters with 20 Ge LDs for a dual readout of heat and light. In this proceeding, we describe the final results of CUPID-Mo. We show CUPID-Mo reached performances compatible with or close to the CUPID goals. We describe the new limits on double beta decays to both ground and excited states. In addition, we present the results of a detailed background model of the experimental data, leading to a measurement of the lowest ever background index for a bolometric $0\nu\beta\beta$ experiment. %Finally, limits on beyond Standard Model processes distorting the spectral shape are presented.
}

\section{Introduction}
\begin{figure}
    \centering
     \includegraphics[height=0.3\textwidth]{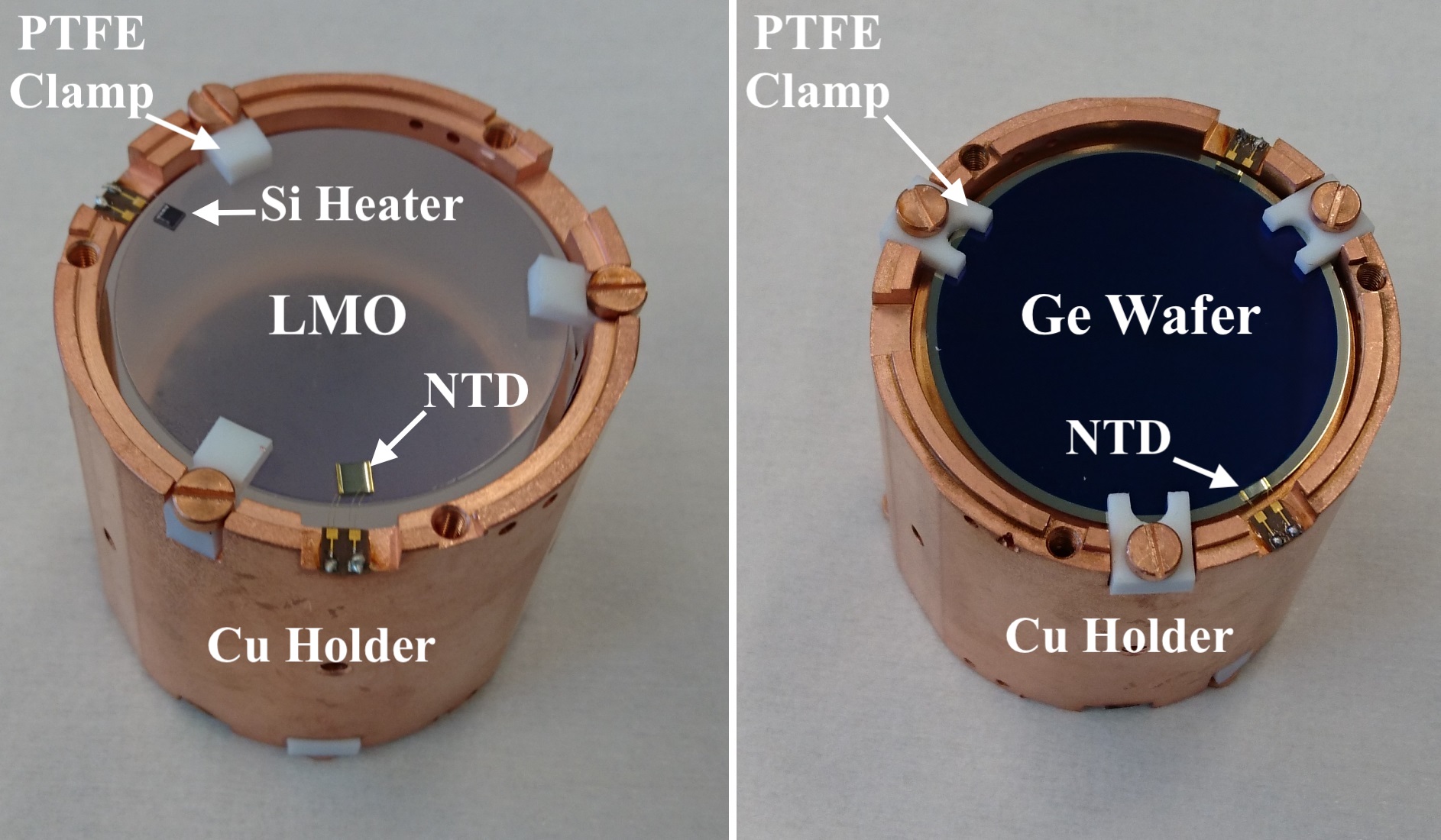}
    \hskip 1cm
            \includegraphics[height=0.3\textwidth]{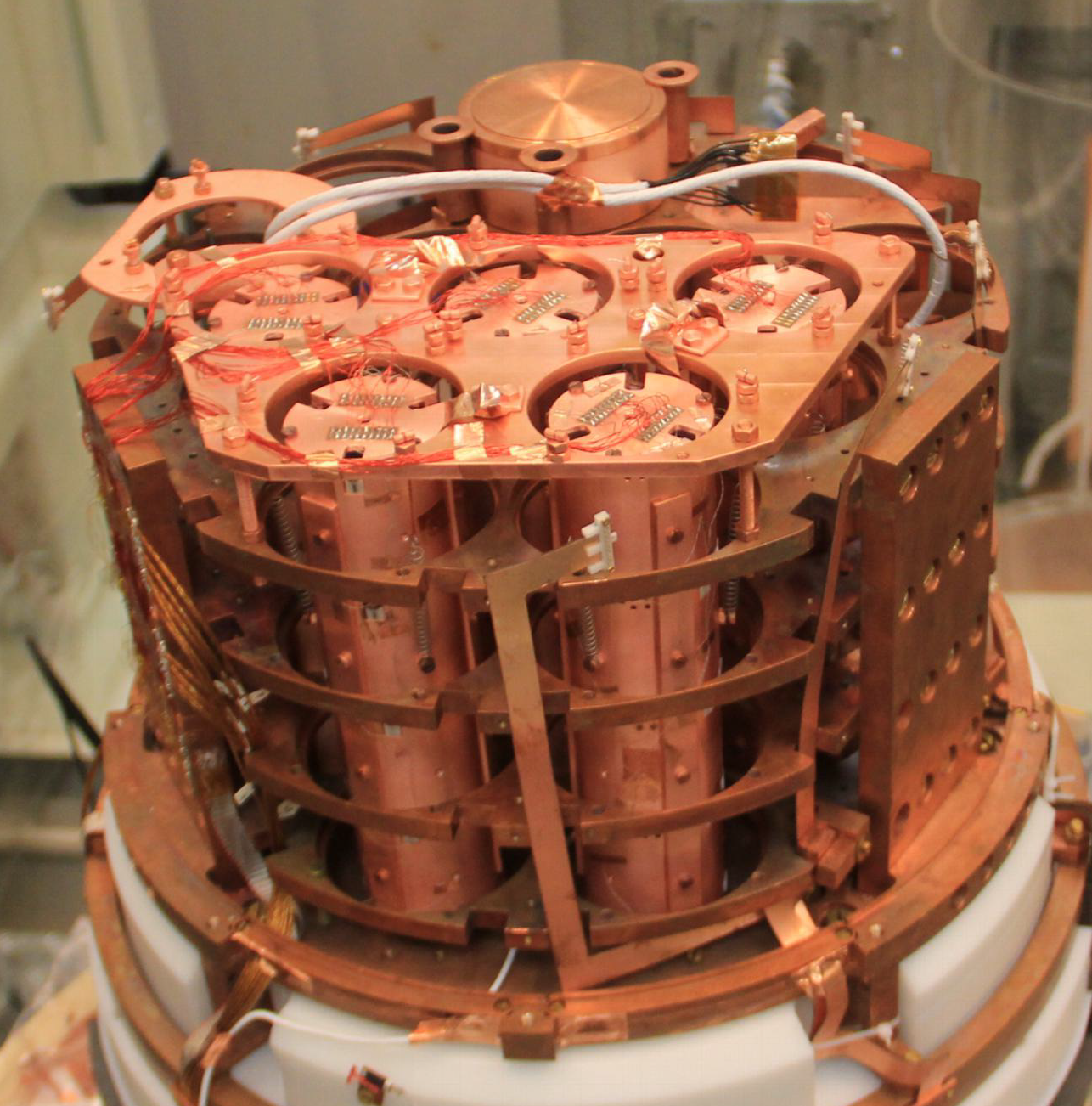}

    \caption{Left: An individual CUPID-Mo detector shown from both the top and bottom, we label the LMO detector, Ge LD, copper holder, PTFE clamps and neutron transmutation doped (NTD) thermistor. Right: The CUPID-Mo detectors mounted in the EDELWEISS cryostat.
    }
    \label{fig:exp}
\end{figure}

Neutrino oscillations show that neutrinos have mass \cite{sno,sk}, however the origin of this mass is still unknown. One way to probe this is with neutrinoless double beta decays ($0\nu\beta\beta)$, or the {\it creation of electrons}\cite{rev}. This process consists of:
\begin{equation}
    (A,Z)\rightarrow(A,Z+2)+2e^{-},
\end{equation}
with the emission of only two electrons and no other particles. Since Lepton number is violated the observation of this process would provide clear evidence of beyond Standard Model physics. The experimental signature would be a mono-energetic peak in the summed electron energy.
\\ \indent One of the most promising experimental techniques to search for $0\nu\beta\beta$ decay is with bolometers (or cryogenic calorimeters). The CUORE experiment \cite{cuore} has stably operated an array of 988 TeO$_2$ bolometers for the past several years. However, the sensitivity to $0\nu\beta\beta$ is limited by a background from $\alpha$ particles. To improve the sensitivity CUPID (CUORE Upgrade with Particle IDentification), will switch from TeO$_2$ to Li$_2$MoO$_4$ (LMO) bolometers and employ a dual readout of heat and scintillation light to remove the $\alpha$ background.
\\ \indent CUPID-Mo was a demonstrator experiment for CUPID \cite{cumo}. It consisted of an array of 20 LMO bolometers and 20 Ge light detectors (LDs) for particle identification. An individual CUPID-Mo module consisting of an LMO bolometer, Ge LD and copper holder is shown in Fig. \ref{fig:exp} (Left). The 20 detector modules were then arranged into 5 towers and mounted in the EDELWEISS cryostat \cite{edel} at the Laboratoire Soutterain de Modane (LSM) as shown in Fig. \ref{fig:exp} (Right).
\\ \indent CUPID-Mo took data between 2019 and 2020 collecting 2.71 kg-yr of exposure of LMO. It achieved performance close to the CUPID goals of:
    \begin{itemize}
        \item  Energy resolution: $\sim 7.4\pm 0.4$ keV FWHM at 3034 keV
        \item Crystal radiopurities: $<0.5\ \mu \mathrm{Bq/kg}$ for $^{228}$Th and $^{226}$Ra
        \item  $\alpha$-particle rejection: $>99.9$\% rejection
        \item Selection efficiency: $\sim 90$ \%.
    \end{itemize}
    These results let to this technology being chosen for the next generation experiment CUPID \cite{cupid}. In addition, CUPID-Mo was an important experiment in its own right with the potential to study the $2\nu\beta\beta$ decay of $^{100}$Mo and set world leading limits on many physics processes.
\\ \indent In this proceeding we will first briefly summarize the CUPID-Mo data analysis in Section \ref{data}. In Section \ref{double} we will describe our extracted limits on double beta decay, both to ground and excited final states. Next we will describe the CUPID-Mo background model in Section \ref{bkg}. We will give the status of measurements of the $2\nu\beta\beta$ decay spectral shape including constraints on beyond Standard Model physics. Finally, in section \ref{pros} we will give the prospects for CUPID.
\section{Data analysis}
\label{data}
\begin{figure}
    \centering
    \includegraphics[width=0.9\textwidth]{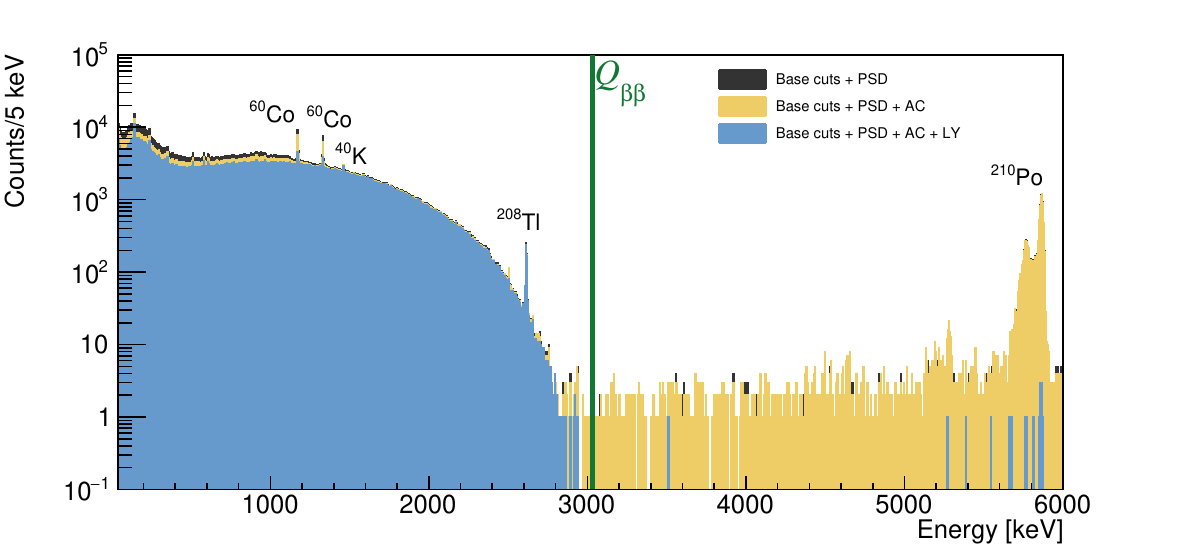}
    \caption{CUPID-Mo experimental data, showing the effect of the pulse shape discrimination (PSD) selection cut, anti-coincidence (AC) cut requiring multiplicity equal to one and light yield cut selecting $\gamma/\beta$ events.}
    \label{cuts}
\end{figure}

For this work we use the full CUPID-Mo data corresponding to $2.71$ kg-yr of LMO. The data were processed using the {\bf DIANA} software developed by the CUORE, CUPID-0 and CUPID-Mo collaborations.
This analysis is described in detail in \cite{cumo_final}.
In particular an optimal filter which maximizes the signal to noise ratio is used to trigger events and evaluate their amplitude. Calibration and thermal gain correction are performed using events from Th/U calibration data. Non-physical events such as pileup are removed using a pulse shape discrimination cut based on principal component analysis \cite{cumo_pca}. A cut on the light detector energies is used to remove $\alpha$ particle induced events. The {\it multiplicity} of an event is defined as the number of triggered detectors with energy above a 40 keV analysis threshold within a $\pm 10$ ms window. Due to the short range of electrons in LMO $2\nu\beta\beta$ and $0\nu\beta\beta$ decay to ground state signal events are likely to reconstruct as multiplicity one ($\mathcal{M}_1$). Multiplicity two ($\mathcal{M}_2$) events are used to constrain backgrounds and extract measurements of double beta decays to excited states. We also remove probable muon events using coincidences with a muon veto system and events likely originating from Th/U decay chains with a delayed coincidence cut (see \cite{cumo_final} for details).
The CUPID-Mo data is shown in Fig. \ref{cuts}, we also show the effect of the data selection cuts. One can see that the light yield cut almost completely removes the background from $\alpha$ particles above 3 MeV.

\section{Limits on double beta decays}
\label{double}
\subsection{$0\nu\beta\beta$ to g.s.}

\begin{figure}
    \centering
    \includegraphics[height=0.3\textwidth]{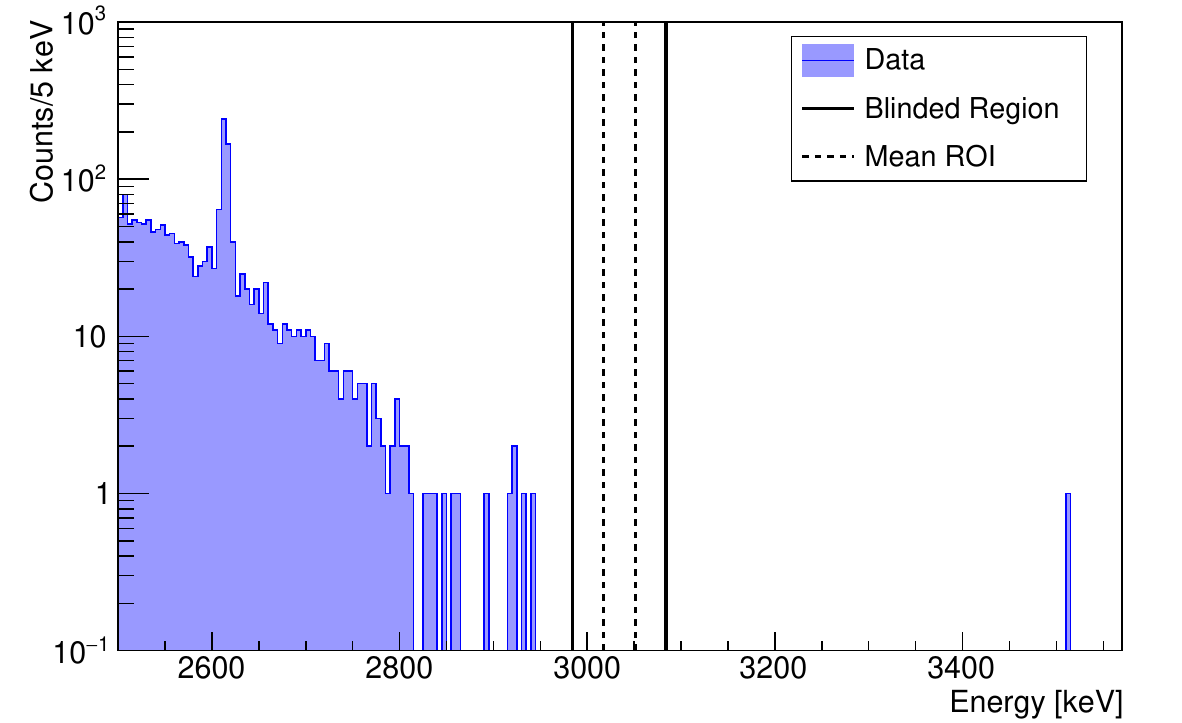}
        \includegraphics[height=0.3\textwidth]{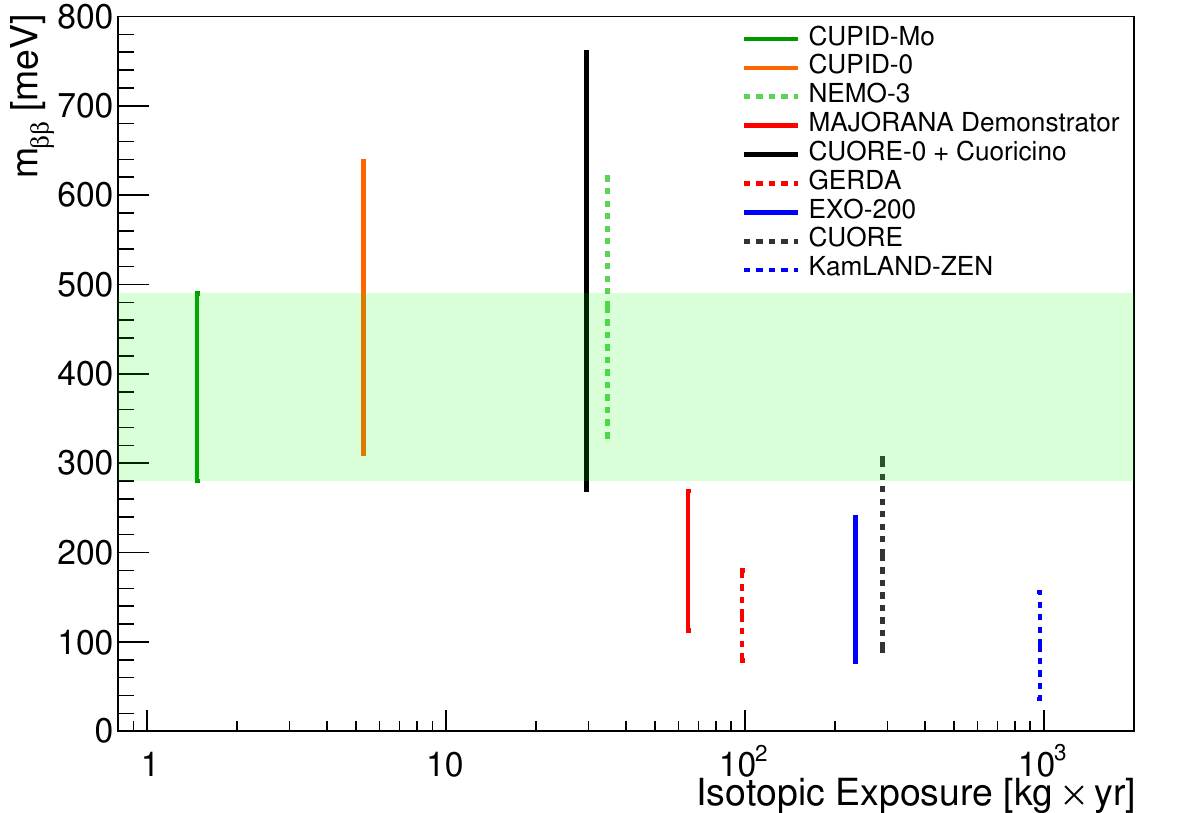}
    \caption{Left: The region of interest (ROI) for the $0\nu\beta\beta$ analysis, the data is shown as a histogram, while the blinded region is shown as a solid line and the mean ROI dashed. No events are observed in the region of interest. Right: The CUPID-Mo limit on $0\nu\beta\beta$ compared to those from other experiments as a function of exposure of target isotope.}
    \label{fig:0v_ROI}
\end{figure}

We perform a Bayesian counting analysis to search for $0\nu\beta\beta$ to the ground state \cite{cumo_final}. After unblinding the data no events were found in the region of interest, as shown in Fig. \ref{fig:0v_ROI} (Left). This leads to a limit of:
\begin{equation}
    T_{1/2}>1.8\times 10^{24} \ \mathrm{yrs}.
\end{equation}
Under the Light Majorana neutrino exchange mechanism a limit of $m_{\beta\beta}<280-490$ meV. This is the most stringent constraint obtained with $^{100}$Mo and is compared to those from other experiments in Fig. \ref{fig:0v_ROI} (Right). CUPID-Mo is competitive with experiments with orders of magnitude larger exposure.
\subsection{Decays to excited states}
\begin{figure}
    \centering
      \includegraphics[height=0.34\textwidth]{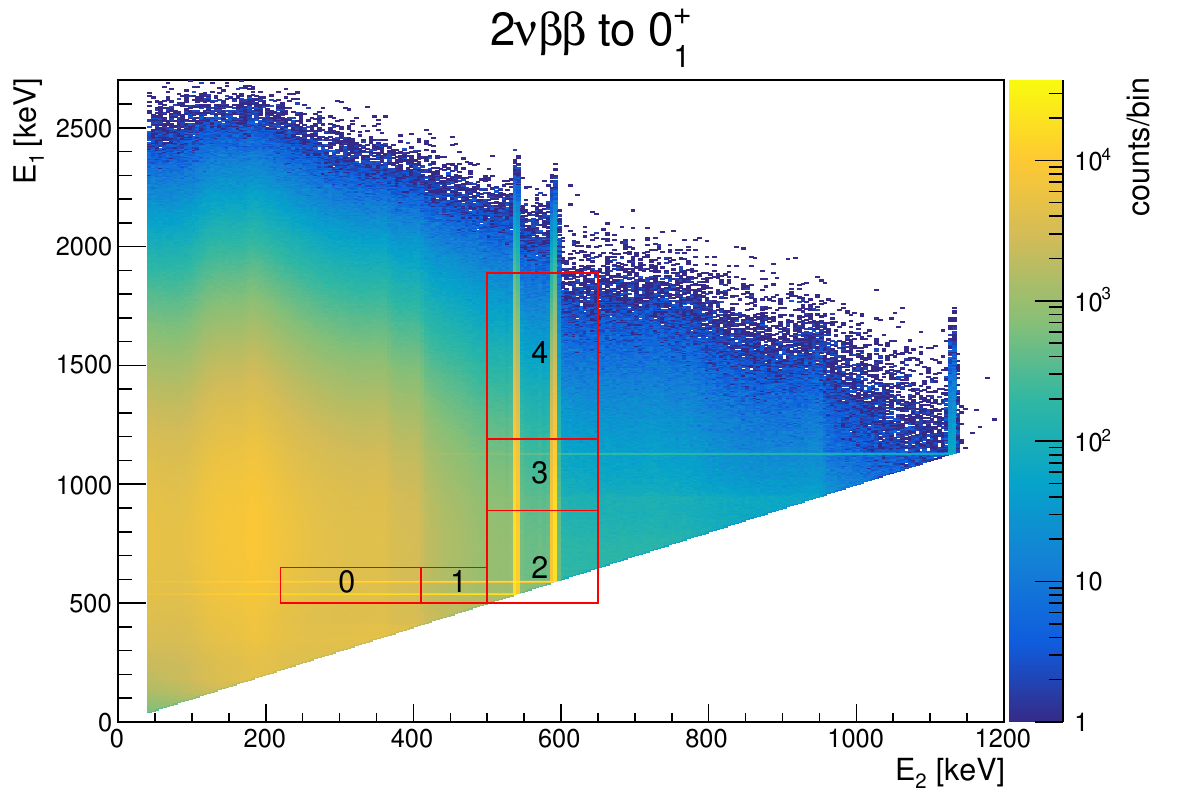}
    \includegraphics[height=0.34\textwidth]{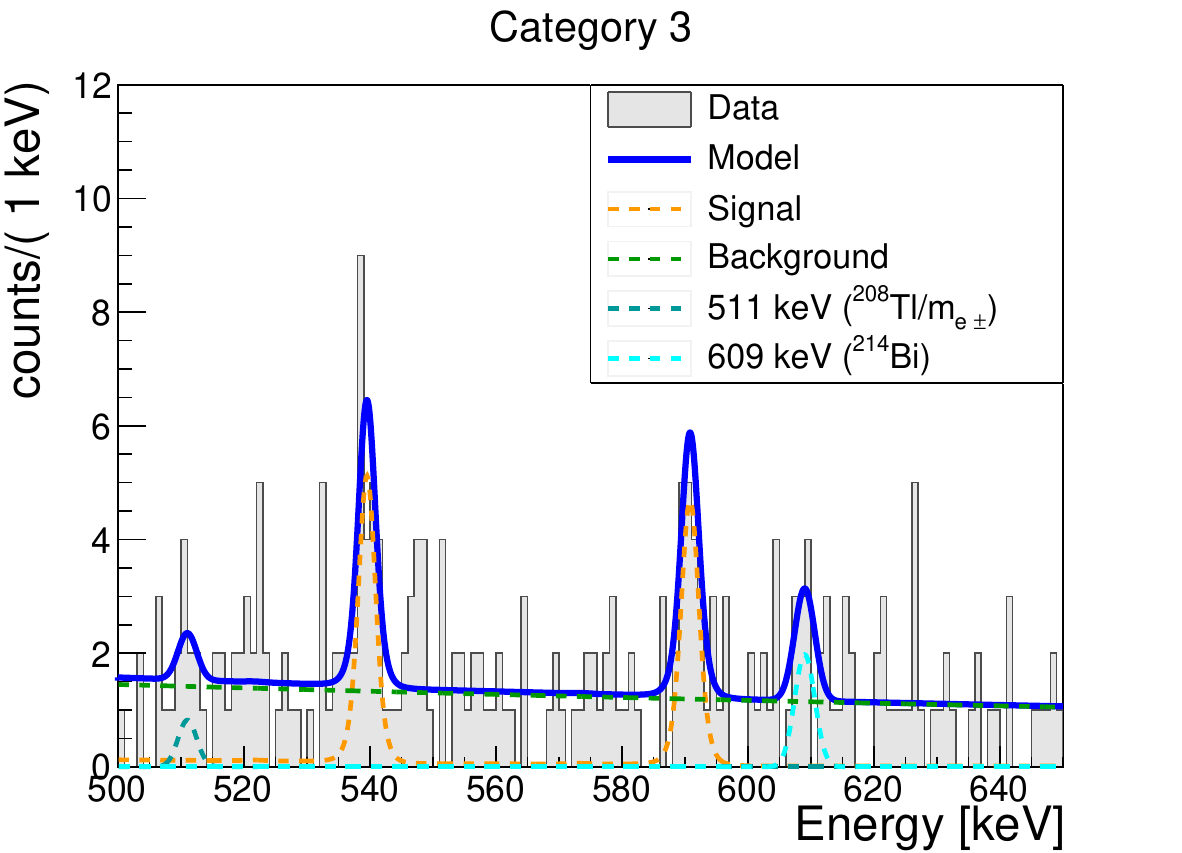}
        
\label{ES_2D}
\caption{Left: Two dimensional energy distribution for $2\nu\beta\beta$ to $0_1^+$ excited state, for $\mathcal{M}_2$ events. We show the categories of events used for the analysis. Right: An example fit to one category showing the fit model consisting of an exponential background and both background and signal $\gamma$ peaks.}
\end{figure}

We also performed a search for double beta decays (both $2\nu\beta\beta$ and $0\nu\beta\beta$) to excited states as described in \cite{es}. $2\nu\beta\beta$ decays to the $0_1^+$ excited state can provide information to constrain nuclear physics models. In addition, both $0\nu\beta\beta$ and $2\nu\beta\beta$ decays to excited states can be sensitive to exotic physics such as Bosonic neutrinos \cite{bosonic}.
\\ \indent In these decays, $\gamma$ particles are produced in coincidence to the $\beta$'s. Due to the longer range of $\gamma$ with respect to $\beta$ particles these events often reconstruct as multi-detector events. We perform a topological search by searching for patterns of energies consistent with being from a $\beta\beta$ decay to excited states. For example as shown in Fig. \ref{ES_2D}, for events with multiplicity two, categories are defined based on the two energies. We then project onto the energy variable containing the $\gamma$ line and perform a simultaneous Bayesian fit, one example fit is shown in Fig. \ref{ES_2D} (Right).
\\ \indent After unblinding the data we observed a clear signal of $2\nu\beta\beta$ to the $0_1^+$ excited state leading to a measurement of the half-life of:
\begin{equation}
      T_{1/2}(2\nu\rightarrow0_1^+)= 7.5\pm 0.8 \ \text{(stat.)} ^{+0.4}_{-0.3} \ \text{(syst.)} \ \times 10^{20} \ \mathrm{yrs}.
\end{equation}
No evidence for $2\nu\beta\beta$ decays to $2_1^+$ excited state or $0\nu\beta\beta$ to $0_1^+$ or $2_1^+$ state was observed leading to limits:
\begin{align}
         T_{1/2}(2\nu\rightarrow2_1^+)&>4.4\times 10^{21} \ \mathrm{yrs} \ (90\% \ c.i.), \\
                        T_{1/2}(0\nu\rightarrow0_1^+)&>1.2\times 10^{23} \ \mathrm{yrs} \ (90\% \ c.i.), \\
            T_{1/2}(0\nu\rightarrow2_1^+)&>2.1\times 10^{23} \ \mathrm{yrs} \ (90\% \ c.i.).
\end{align}
These are the most stringent limits on these processes for $^{100}$Mo.

\section{Background model }
\label{bkg}
\begin{figure}
    \centering
    \includegraphics[width=0.7\textwidth]{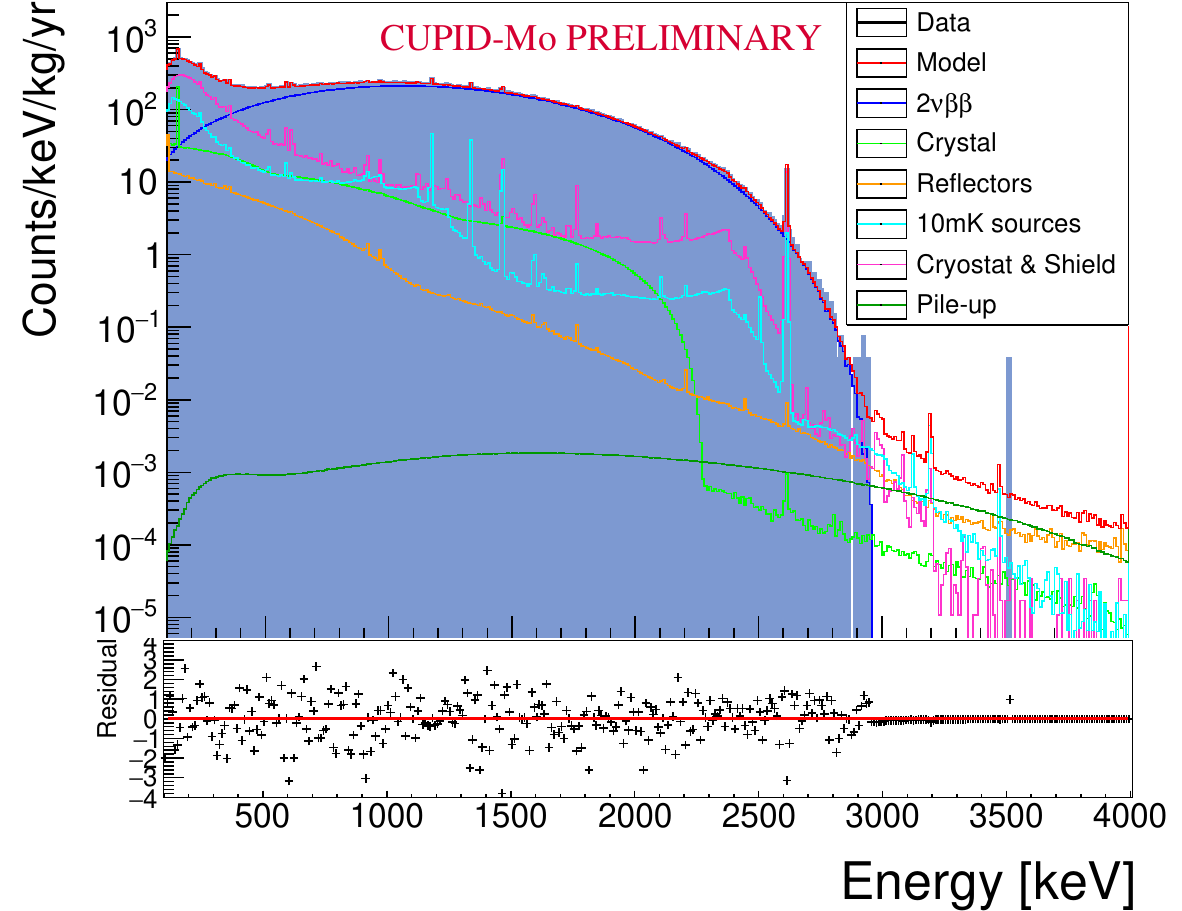}
    \caption{The background model reconstruction of the $\mathcal{M}_{1,\gamma/\beta}$ data, we chose the fit and the various contributions to the model in the upper panel. In the lower panel we show the residual defined as ~$(N_\text{Data}-N_\text{Fit})/\sqrt{N_\text{Data}}$.}
    \label{best_fit}
\end{figure}

 The CUPID-Mo background suppression leads to a very clean $2\nu\beta\beta$ spectrum. A detailed understanding of the experimental backgrounds \cite{bkg} is important both to exploit this spectrum for physics measurements and to provide information on the radioactivity of components for CUPID.
 \\ \indent We perform Monte-Carlo simulations using Geant4 of the various contributions to the background. These can be categorized into:
 \begin{itemize}\item  $2\nu\beta\beta$ signal, 
 \item radioactivity in the LMO crystals, \item contamination in the reflective foils and other materials surrounding the detectors, 
 \item sources inside the cryostat 10 mK vessel,
 \item contamination in the cryostat shields, 
 \item pileup events where two energy deposits occur close enough in time to be indistinguishable from a single event.
 \end{itemize}
 We perform a Bayesian fit to our experimental data over 3 data spectra: $\mathcal{M}_{1,\gamma/\beta}$ (single crystal events tagged as $\gamma$ or $\beta$), $\mathcal{M}_{1,\alpha}$ (single crystal events at high energy likely to be from $\alpha$ particles) and $\mathcal{M}_2$ events (two crystal events where we fit the summed energy). This fit is implemented using Markov Chain Monte Carlo with JAGS \cite{jags}.
 \\ \indent We show the best fit reproduction of the $\mathcal{M}_{1,\gamma/\beta}$ data in Fig. \ref{best_fit}. One can see that the model is able to describe very well the experimental data and that the $2\nu\beta\beta$ decay signal to background ratio is very high. We reconstruct the background index in the region of interest as:
 \begin{equation}
     b=3.8\pm 1.0\times 10^{-3} \ \text{cts/}\text{FWHM}/\text{mol}_{\text{iso}}/\text{yr},
 \end{equation}
this is the lowest background index in a bolometric double beta decay experiment.
\\ \indent
 This model also enables us to extract the radioactivity of the LMO detectors which is consistent with the CUPID requirements (see \cite{bkg}).

\subsection{$2\nu\beta\beta$ decay measurement}
The background model reconstructs the half-life of $2\nu\beta\beta$ decay to the ground state. We consider systematic uncertainties related to the number of $^{100}$Mo nuclei, energy reconstruction, theoretical spectral shape, binning, model choice and selection efficiencies. The analysis is currently being finalized and will be presented in a future publication.

%We extract a preliminary measurement of $T_{1/2}$ of:
%\begin{equation}
 %   T_{1/2} = 6.99\pm 0.02 \ \text{(stat.)}\ ^{+0.12}_{-0.10} \text{(syst.)} \ \times 10^{18} \ \mathrm{yrs}.
%\end{equation}
%This is the most precise measurement of a double beta decay to ground state in any isotope.
\subsection{Exotic physics processes}
 Beyond Standard Model (BSM) physics processes can distort the $2\nu\beta\beta$ spectrum \cite{nemo}.
 The spectral shape of Standard Model $2\nu\beta\beta$ decay is proportional to $(Q_{\beta\beta}-E)^n$ where $n=5$ is the {\it spectral index} of the decay. Beyond Standard Model spectra can depend on a different spectral index.
\\ \indent
 We are performing a search for $2\nu\beta\beta$ decay with Lorentz Violation (LV) and $0\nu\beta\beta$ with the emission of Majorons.  We insert one by one the BSM spectral shapes into the background model fit. The analysis is being finalized and we expect to present results in a future publication.
 % No evidence for any of the processes is found. We extract preliminary limits of $0\nu\beta\beta$ decay with the emission of Majorons shown in Tab. \ref{maj}.
 %\\ \indent LV is parameterised by $a_{of}^{(3)}=C \times \Gamma_{LV}/\Gamma_{SM}$. We extract a preliminary limit on this parameter of:
 %\begin{equation}
 %    a_{of}^{(3)}<2.2 \times 10^{-6} \ \mathrm{GeV} \ \text{(90 \% c.i.}.
 %\end{equation}
%This limit is compared to constraints from other experiments in Fig. \ref{cpt}. We observe that the constrain from CUPID-Mo is the second strongest, only worse than that of the NEMO-3 experiment. However NEMO-3 had a much larger isotope mass and used a different technique measuring both the single and summed electron energies.
 %\begin{table}[]
  %          \centering
   %         \caption{Exclusion limits on $0\beta\beta$ decay with one of more Majorons.}
    %        \begin{tabular}{c|c}
     %        Process &    Limit [$10^{21}$ yrs] (90\% c.i.)\\
      %         \hline $ \beta\beta\chi_0$ (n=1) & 2.1 \\
       %        $\beta\beta\chi_0$ (n=2) & 4.5 \\
        %       $\beta\beta\chi_0(\chi_0)$ (n=3) & 1.4 \\
         %      $\beta\beta\chi_0\chi_0$ (n=7) & 0.5
          %  \end{tabular}
     %   \end{table}
      %  \label{maj}
%        \begin{figure}
 %          \centering
  %         \includegraphics[width=0.6\textwidth]{Maj-3.pdf}
   %       \caption{Spectral shapes for various BSM processes distorting the $2\nu\beta\beta$ decay spectral shape, superimposed over the CUPID-Mo data.}
    %    \label{cpt}
   %    \end{figure}
\section{Conclusion}
\label{pros}
The final results of CUPID-Mo confirm that scintillating bolometers based on LMO are a mature technology for CUPID with performance close to the CUPID goals. In addition, we have performed a variety of physics studies with world leading results, this demonstrates the possible sensitivity of CUPID which will have a factor $\sim 100$ $\times$ larger mass and will have a sensitivity to CUPID fully covering the inverted ordering mass region.
\section*{Acknowledgments}

This work has been performed in the framework of the CUPID-1 (ANR-21-CE31-0014) and LUMINEU programs, funded by the Agence Nationale de la Recherche (ANR, France).
We acknowledge also the support of the P2IO LabEx (ANR-10-LABX0038) in the framework ''Investissements d'Avenir'' (ANR-11-IDEX-0003-01 – Project ''BSM-nu'') managed by ANR, France.

The help of the technical staff of the Laboratoire Souterrain de Modane and of the other participant laboratories is gratefully acknowledged. 
We thank the mechanical workshops of LAL (now IJCLab) for the detector holders fabrication and CEA/SPEC for their valuable contribution in the detector conception. 
F.A. Danevich, V.V. Kobychev, V.I. Tretyak and M.M. Zarytskyy were supported in part by the National Research Foundation of Ukraine Grant No. 2020.02/0011. O.G. Polischuk was supported in part by the project “Investigations of rare nuclear processes” of the program of the National Academy of Sciences of Ukraine “Laboratory of young scientists”. A.S. Barabash, S.I. Konovalov, I.M. Makarov, V.N. Shlegel and V.I. Umatov were supported by the Russian Science Foundation under grant No. 18-12-00003. J. Kotila is supported by Academy of Finland (Grant Nos. 314733, 320062, 345869).
Additionally the work is supported by the Istituto Nazionale di Fisica Nucleare (INFN) and by the EU Horizon2020 research and innovation program under the Marie Sklodowska-Curie Grant Agreement No. 754496. This work is also based on support by the US Department of Energy (DOE) Office of Science under Contract Nos. DE-AC02-05CH11231, and by the DOE Office of Science, Office of Nuclear Physics under Contract Nos. DE-FG02-08ER41551, DE-SC0011091; by the France-Berkeley Fund, the MISTI-France fund and  by the Chateau-briand Fellowship of the Office for Science \& Technology of the Embassy of France in the United States. This research used resources of the National Energy Research Scientific Computing Center (NERSC) and the IN2P3 Computing Centre.
This work makes use of the DIANA data analysis software and the background model based on JAGS,  developed by the CUORICINO, CUORE, LUCIFER, and CUPID-0 Collaborations.

\end{document}